\documentclass[jgrga]{AGUTeX}

  \usepackage[dvips]{graphicx}
\authorrunninghead{Li et al}

\titlerunninghead{STUDY OF THE NWC ELECTRONS BELT OBSERVED ON DEMETER SATELLITE}

\begin{document}

%% ------------------------------------------------------------------------ %%
%
%  TITLE
%
%% ------------------------------------------------------------------------ %%

\title{Study of the NWC electrons belt observed on DEMETER Satellite}
%
% e.g., \title{Terrestrial ring current:
% Origin, formation, and decay $\alpha\beta\Gamma\Delta$}
% You may use \\ to break the title over several lines.

%% ------------------------------------------------------------------------ %%
%
%  AUTHORS AND AFFILIATIONS
%
%% ------------------------------------------------------------------------ %%

%Use \author{\altaffilmark{}} and \altaffiltext{}

% \altaffilmark will produce footnote;
% matching altaffiltext will appear at bottom of page.
% May use \\ to start a new line.
\authors{Xinqiao Li\altaffilmark{1}, Yuqian Ma\altaffilmark{1}, Ping
Wang\altaffilmark{1},
 Huanyu Wang\altaffilmark{1}, Hong Lu\altaffilmark{1}, Xuemin Zhang\altaffilmark{2}, Jianping Huang\altaffilmark{2}, Feng
 Shi\altaffilmark{4}, Xiaoxia Yu\altaffilmark{1}, Yanbing
 Xu\altaffilmark{1}, Xiangcheng Meng\altaffilmark{1}, Hui
 Wang\altaffilmark{1}, Xiaoyun Zhao\altaffilmark{1} and M. Parrot\altaffilmark{3}
 }

\altaffiltext{1}{Institute of High Energy Physics, Chinese Academy
of Sciences, Beijing, China. (mayq.ihep@gmail.com,
lixq@ihep.ac.cn)}

\altaffiltext{2}{Institute of Earthquake Science,  China
Earthquake Administration, Beijing, China.}

\altaffiltext{3}{Laboratory of Physics and Chemistry of
Environment and Space , CNRS, Orleans, France}

%% ------------------------------------------------------------------------ %%
%
%  ABSTRACT
%
%% ------------------------------------------------------------------------ %%

% >> Do NOT include any \begin...\end commands within
% >> the body of the abstract.

\begin{abstract}
We analyzed observation data collected by the Instrument for the
Detection of Particles (IDP) on board of DEMETER satellite during
the period of total seventeen months in 2007 and 2008. In the
meantime, the VLF transmitter located at NWC ground station was
shutdown for seven months and working for total ten months. Our
analysis£¬for the first time£¬ revealed in details the transient
properties of the space electrons induced by the man-made VLF wave
emitted by the transmitter at NWC. First, we mapped the electron
flux distribution and figured out the special range what the NWC
belt covered. Then we investigated the NWC electron spectrograms
in a wide range of McIlwain parameter (up to L$\sim$3.0). Finally,
we obtained the averaged energy spectrum of the NWC electrons
within the drift loss-cone, and compared the difference during
  the observations between daytime and nighttime. Our results proved the fact that the VLF emissions
  from NWC transmitter created momentary electron enhancement with fluxes up to 3 orders of magnitude.
  These electrons are distributed in the region of $180^{\circ}$ in longitude and 1.6 $\sim$ 1.9 of L shell.
  In addition, the VLF emission induced either enhancement or loss of electrons in higher magnetic shells
  up to L$\sim$3, and the maximum loss was up to 60$\%$ of the original value. The energy spectra of these
  electrons revealed that the enhancement during the NWC daytime are more attenuated than those in the NWC
  nighttime£¬and that the shape and the cutoff energy of the spectra are also quite different. We will present
  the results of our analysis, compare it with previous studies, and discuss the agreement of our results with
  the theory of wave-particle interaction.

\end{abstract}

%% ------------------------------------------------------------------------ %%
%
%  BEGIN ARTICLE
%
%% ------------------------------------------------------------------------ %%

% The body of the article must start with a \begin{article} command
%
% \end{article} must follow the references section, before the figures
%  and tables.

\begin{article}

%% ------------------------------------------------------------------------ %%
%
%  TEXT
%
%% ------------------------------------------------------------------------ %%

\section{Introduction}
More than thirty years ago, it has been learned that the VLF wave
emitted by transmitters on the ground stations can cause
precipitation of electrons in the radiation belt, and the process
obeys the theory of wave-particle interaction. Subsequently, there
have been numerous experimental observations and theoretical
interpretations on this subject. Bullough et al.(1976) studied the
ELF/VLF radiation at the altitude of the Ariel satellite and
revealed the phenomenon of wave-particle interaction in a
geomagnetic conjugate region in the southern hemisphere(2 $<L<$
3). Kimura et al. (1983)found strong correlations between the
0.3-6.9keV electron fluxes observed by EXOS-B satellite and the
0.3-9kHz VLF wave emitted by transmitter at SIPLE ground station.

In the SEEP experiment conducted from May to September in 1982, by
turning the transmitter at the NAA ground station on and off at
the time scale of seconds, Imhof et al (1983) studied the
instantaneous correlation between the VLF signal emitted from the
ground station and the electron fluxes observed by the
low-orbiting satellite. Inan et al. (1985) interpreted the
observation of the SEEP experiment using the theory of
particle-wave interaction. They proposed that the precipitation of
particles was restricted by pitch angle distribution of those
particles close to the loss cone.

During the same period, William et al. (1993) studied the VLF wave
propagation in the D layer of the earth ionosphere, they proposed
a multiple-mode three-dimensional model. Abel and Thorne (1998)
proposed that in the inner radiation belt the electron flux
reduction mainly induced by the VLF wave scattering, which cause
the electrons fall into the loss cone. Their theoretical
calculations indicated that the energy of wave-particle resonance
decreases with higher L values. Furthermore, Richard et al.(2005)
studied the mechanism of electron acceleration in the outer
radiation belt. Their studies indicate that these electrons could
be accelerated to MeV energy region by VLF wave with several kHz,
and the electron fluxes in the observed region could rise to 3
orders of magnitude within one to two days.

The DEMETER satellite (Detection of Electro-Magnetic Emissions
Transmitted from Earthquake Regions) launched by France in June
2004 is a low-orbiting electro-magnetic satellite with detectors
onboard to measure space electric fields, magnetic fields and high
energy particles and so on (~\cite{Sauvaud2006};
~\cite{Berthelier2006};~\cite{Parrot2006}). These detectors
provide us with the opportunity to further study the interactions
between the man-made VLF wave and particles in the radiation belt.

By turning on and off the transmitter based on the NPM ground
station, Inan, Graf et al studied the correlation between the
ground VLF wave and the electron fluxes detected by on-board IDP
(Instrument for Detecting Particles). The experiments were
performed in ten different time scales ranging from second to
minute, and they found correlations in 0.1 and 0.2 Hz ON/OFF
frequency (~\cite{Inan2007}). However, subsequent experiments
using the 0.1 Hz frequency (i.e. on for 5 sec and off for 5 sec,
alternately ) indicated that the correlation rate is only 13.9$\%$
(~\cite{Graf2009}). Quite recently P. Wang et al.(2010) pointed
out, the lower correlation rate may be caused by that IDP only can
measure the electrons with big pitch angles.

Parrot, Sauvaud,  Gamble, et al. studied the relation between the
VLF emission from the transmitter at NWC ground station and the
observation by IDP in relatively larger regions in space and
longer time intervals. There were very strong ionosphere
disturbance detected near the NWC ground station
(~\cite{Parrot2007}). Moreover, due to the VLF wave emitted by the
NWC transmitter, plenty electrons precipitate into the drift loss
cone and drift eastward until into the SAA (South Atlantic
Anomaly) to covered a region with L value between 1.4~1.7 during
the local night period. The study also reported that the spectrum
of local electrons has a wing-like structure (~\cite{Sauvaud2008};
 ~\cite{Gamble2008}). There are also more relevant studies,
for example, the numerical simulation for precipitation
characteristics induced by five ground-based VLF transmitters
(~\cite{Kulkarni2008}), the restriction of VLF wave in (18-25 kHz)
in driving electron precipitation in inner radiation belt, as well
as the effect of NAA station to electrons in outer radiation belt
(~\cite{Clilverd2008};~\cite{Clilverd2010}).

Furthermore, recent studied have derived the theoretical models of
the NWC VLF wave transmission in the ionosphere
(~\cite{Lehtinen2009}), and the different effect by ducted or
non-ducted VLF wave propagation (~\cite{Rodger2010}). A report
revealed the increasing of the MF component at the height of the
DEMETER orbit over the VLF transmitters (~\cite{Parrot2009}). It
is caused by the emissions of the global thunderstorm activity
which are less attenuated when they cross the ionosphere above
these locations.

We analyzed the IDP data during 2007 and 2008, when the NWC
transmitter was turned off from June 2007 to February 2008 and
turned on in the rest of the time. NWC transmitter has strong
radiation power and the very narrow bandwidth of the VLF
emission£¬it provided us with the best opportunities in studying
the mechanism of wave-particle interactions. In
particular£¬because the NWC station was kept either on or off for
a long period, we have enough data to make a comprehensive study
on the effect of man-made electron precipitation at high
statistical significant level.

We analyzed the IDP data during 2007 and 2008, when the NWC
transmitter was turned off from June 2007 to February 2008 and
turned on in the rest of the time. NWC transmitter has strong
radiation power and the very narrow bandwidth of the VLF
emission£¬it provided us with the best opportunities in studying
the mechanism of wave-particle interactions. In
particular£¬because the NWC station was kept either on or off for
a long period, we have enough data to make a comprehensive study
on the effect of man-made electron precipitation at high
statistical significant level.

For the first time, we analyzed the data observed by IDP during 7
months of NWC off and 10 months of NWC on. By using the (ON - OFF)
method, we mapped the instantaneous electron belts, which are
induced by the VLF wave emitted by NWC transmitter and are termed
"NWC electron belt", or "NWC belt" in brief. We also
quantitatively studied the time-averaged characteristics of the
NWC electron belts, for which we will present the distribution
range along the longitude and magnetic shell of L parameter, the
time-averaged energy spectrum of the precipitation electrons, the
variation of the energy spectrum in different L shell intervals,
and the diurnal variation of the electrons energy spectrum. The
results of the analysis will be presented and discussed.

%\section{Description of Experiment}

\section{Statistical analysis of the NWC electron flux}
The orbit of the DEMETER satellite is at the height of 670km
(after 2006). It is a quasi-sun-synchronous orbit, and the orbital
inclination is 98¡ã. The satellite is three-axis-stabilized.
During local daytime the satellite flies downward (from north to
south) and the descending node is 10:30 at noon; during local
nighttime it flies upward (from south to north) and the ascending
node is 22:15 at midnight. The orbital period is 102.86 minutes.
The repeated period of orbit location is around half of a month.

The NWC ground station locates in the northwest corner of
Australia at (21.82 $^{\circ}$S, 114.15 $^{\circ}$E). Its
geomagnetic coordinate is (31.96 $^{\circ}$, 186.4 $^{\circ}$).The
NWC transmitter works on 19.8 kHz waveband with very narrow
bandwidth of 300Hz, and the great emission power of 1MW.

\begin{figure}
\center{
 \noindent\includegraphics[width=8pc]{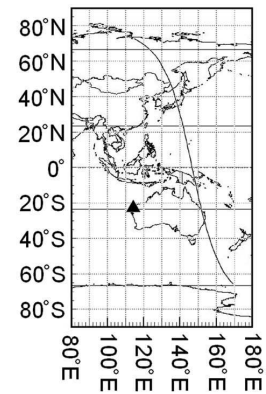}
 \caption{The location of NWC station and the trajectory of the orbit No.20520 (upward) \label{world}
 }
}
 \end{figure}

\begin{figure}
\center{
 \noindent\includegraphics[width=18pc]{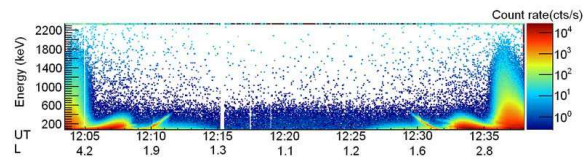}
 \caption{Electron spectrogram versus UT/LT and L shell of orbit No. 20520 (upward) \label{spec20520}
 }
}
 \end{figure}

The IDP is placed on flank of the DEMETER satellite with field of
view (FOV) of 32$^{\circ}$, which was pointing towards the west
during the satellite upward flying at night and towards the east
during the satellite downward flying at day time. So in the most
case, only those electrons with big pitch angle close to 90 degree
can be detected by IDP. Figure \ref{world} shows the location of
NWC station and a trajectory of a satellite orbit in east side.
Figure \ref{spec20520} presents electron spectrogram detected by
IDP along with this orbit. It should be noted that in Figure
\ref{spec20520} there are two wing-like spectrum structures at
lower energy band around L$\sim$1.9 in the north and L$\sim$1.6 in
the south, which correspond to the precipitated electron belt
induced by the NWC VLF wave. These will be the focus of this
study.

\begin{figure*}
\center{
\noindent\includegraphics[height=.75\mycolumnwidth,width=15pc]{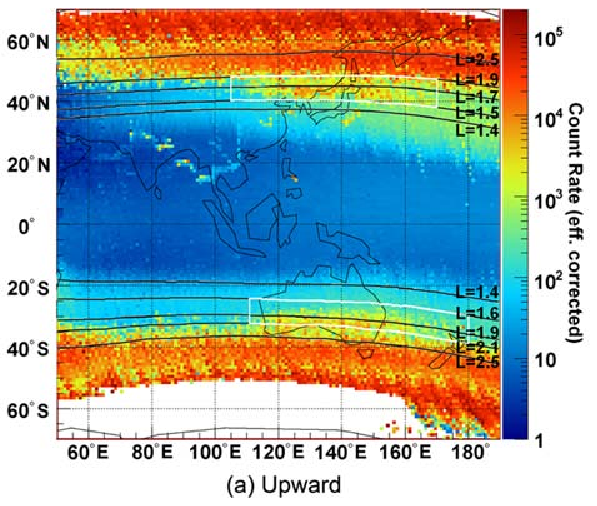}
\noindent\includegraphics[height=.75\mycolumnwidth,width=15pc]{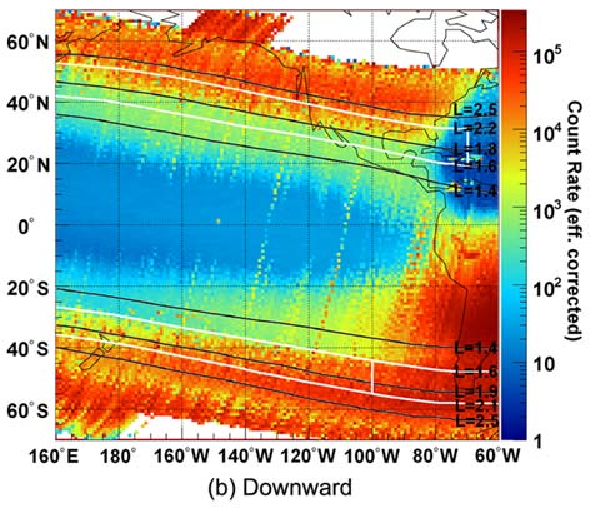}
\caption{The electron flux distribution with the energy range
108-411keV for observations in 2008. Colors indicate the average
flux value within each of (1$^{\circ}$,1$^{\circ}$) pixel. The
simplified unit of counting rate(counts/s) is obtained by
multiplying the factor of 1.16(cm$^{2}\cdot$sr) with the averaged
flux value. \label{updownmaps}} }
\end{figure*}

We mapped the electron flux with the data in the entire year 2008.
The flux was averaged in each of the 1$^{\circ}$by 1$^{\circ}$
pixels along the geographic latitude and longitude. The result was
shown in Figure \ref{updownmaps}. The white boxes are the main
regions of the VLF man-made electron belts corresponding to the
"wing" structure shown in Figure 1b). The black lines mark the
value of the McIlwain parameter L at the height of the DEMETER
satellite. Figure \ref{updownmaps}(a) shows the range of the VLF
electrons "wing" structure detected during the upward flying.
Whereas during the downward flying, the VLF electron belts were
detected to move toward east, extending all the way to the South
Atlantic Anomaly (SAA) region in the southern hemisphere, as shown
in Figure \ref{updownmaps}(b).

The two white boxed regions in Figure \ref{updownmaps} (a) (upward
data) correspond to the electrons belt near NWC position (SL:
South Local) and the conjugation position of the NWC station (NL:
North Local); whereas the three white regions in Figure
\ref{updownmaps} (b) (downward data) correspond to the SE (South
East) region and NE (North East) region as well as the part of SE
region extending into SAA region, which is belong to SE and marked
separately.

Using upward orbit data, the same mapping have been made for each
of individual month in 2007 and 2008. We find that the electrons
precipitation belts in NL and SL region vanished from June 2007 to
January 2008. So that the statistical analysis have been made for
each of 24 months and both of upwards and downwards data
respectively.

We select the longitude range of $\lambda$ in (110$^{\circ}$,
170$^{\circ}$) for NL and SL regions£¬£¨(170$^{\circ}$,
260$^{\circ}$)for NE and SE regions, the L value in (1.6, 1.8)
range, and the electron energy range of 206$\pm$26.7keV for the
analysis. The average count rate are calculated in each of
($\Delta$L, $\Delta\lambda$) with unit size of (0.01, 1$^\circ$)
and accumulated to get flux distributions within the selected
region.

\begin{figure}
\center{
\noindent\includegraphics[width=13pc]{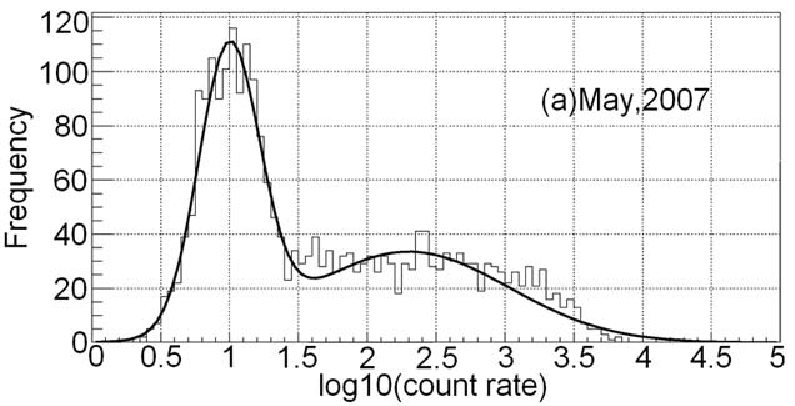}
\noindent\includegraphics[width=13pc]{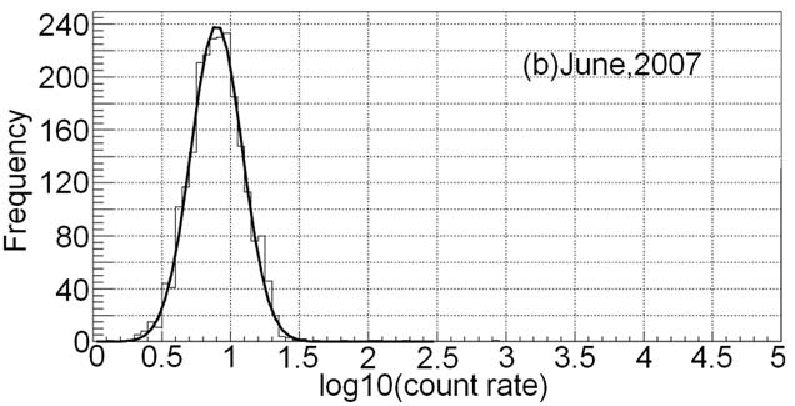}
\caption{The distribution of average electrons flux along
logarithm (NL and SL). Energy range: 206$\pm$26.7keV.
\label{meancurve}} }
\end{figure}

As examples, Figure \ref{meancurve} displays the result with the
data in May and June 2007 respectively, which can be fit by
(a)double Gaussian distribution, and (b) single Gaussian
distribution in logarithm scale of averaged counting rate.

Compared with June, the distribution in May has one more component
with very wide range of higher count rate, and the maximum value
can be more than 1000 times of the average background value.
Figure 5 shows the statistics result for each of 24 months in the
two years. We can find that it has two components: a background
component (mean1$\pm$sigma1) with constant low count rate and a
variable high count rate component (mean2$\pm$sigma2).

This high count rate component was distributed in a wide range and
more variable. It vanished from June 2007 to January 2008 which is
consistent with the NWC off time, i.e. the time when there was no
signal of the NWC transmission as mentioned in the paper by Zhang
et al. (2009). So we can confirm that the high count rate
component is contributed by the NWC transmission, because its
appearance and disappearance clearly synchronize with the NWC on
and off.

\begin{figure}
\center{
\noindent\includegraphics[width=13pc]{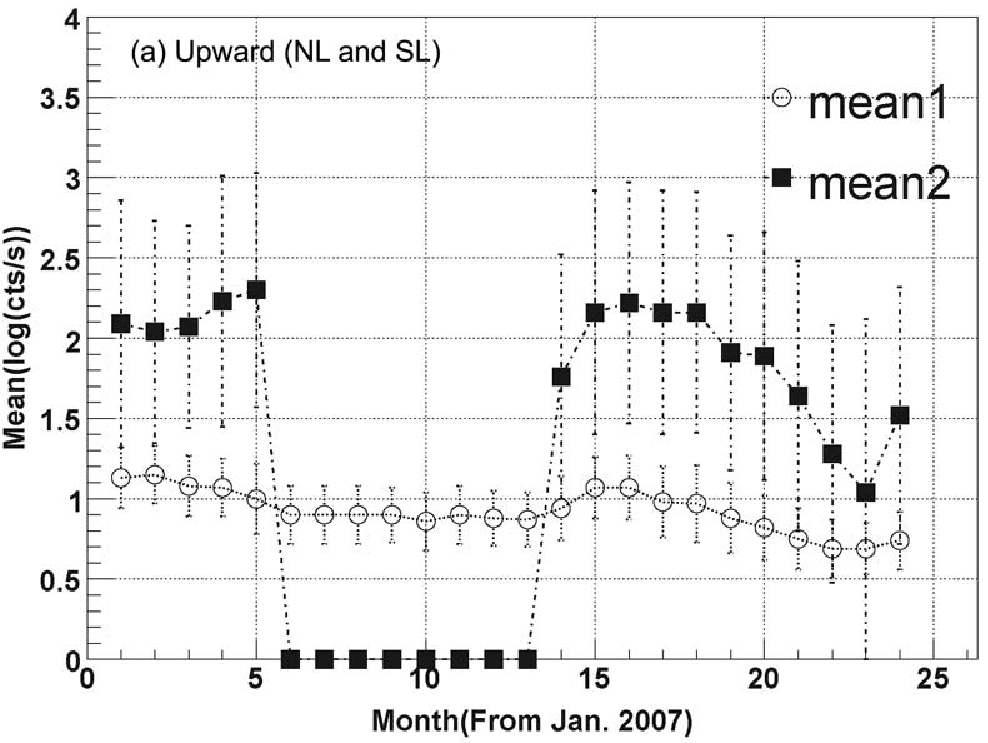}
\noindent\includegraphics[width=13pc]{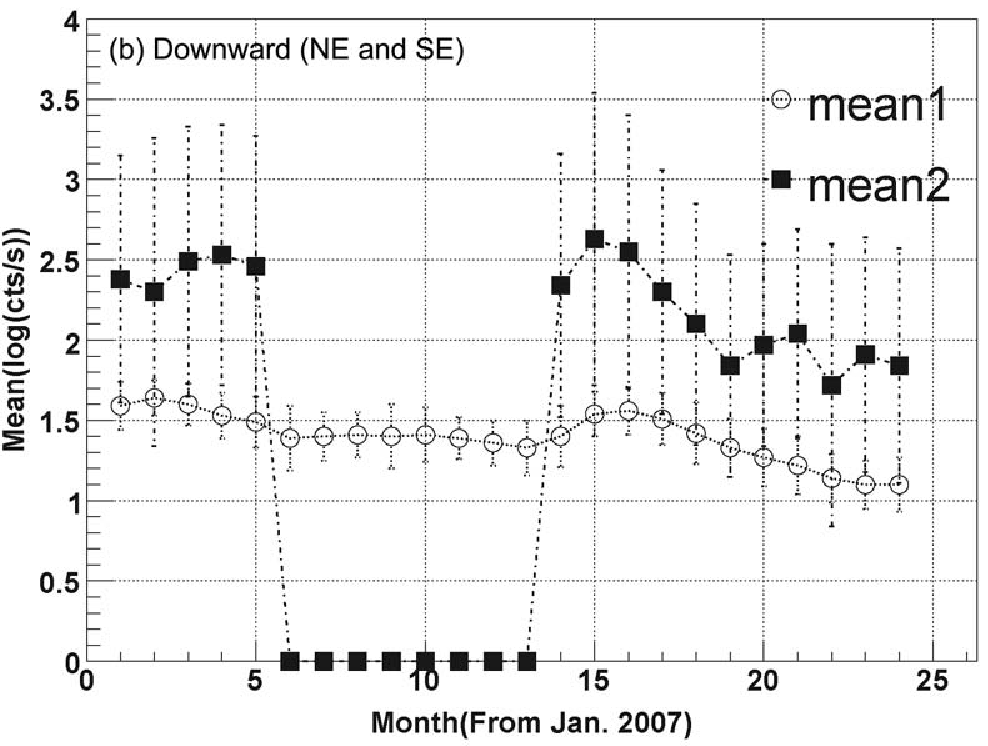}
\caption{The mean and sigma of double Gaussian fit for each months
of 2007 and 2008. Circles represent the first component and the
black squares represents the second
component.\label{meansigmacurve}} }
\end{figure}

Based on the ON and OFF status shown in Figure
\ref{meansigmacurve}, we selected "ON data" and "OFF data",
respectively. The ON data include the observations in ten months
from March to May in 2007 and March to September in 2008, while
the OFF data include the observations in seven months from July
2007 to January 2008. Since the Solar activity was keeping
relatively quiet during 2007 and 2008, by subtracting the
background "OFF data" from the "ON data", (ON - OFF), we can
smooth out short term random fluctuations. In addition, we can
conduct detailed studies on the basic dynamic characteristics of
the NWC electron belts using large data sets to get better
statistical significance. Since the electron flux varies every
month and NWC electrons are moving all the time, the obtained
results reflect the average statistical characteristics. The
results will be given in the following sections.

\section{Spatial distribution of NWC electrons}\label{sec3}
\begin{figure}
\center{ \noindent\includegraphics[width=8pc]{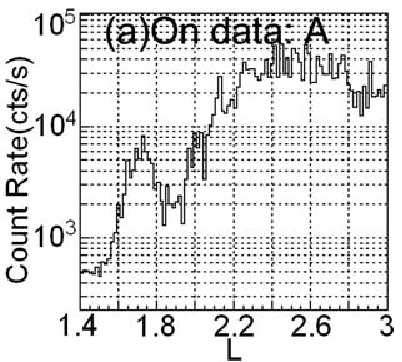}
\noindent\includegraphics[width=8pc]{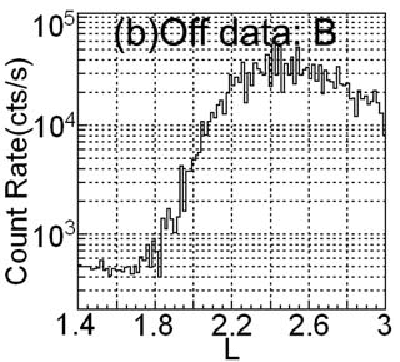}
\noindent\includegraphics[width=8pc]{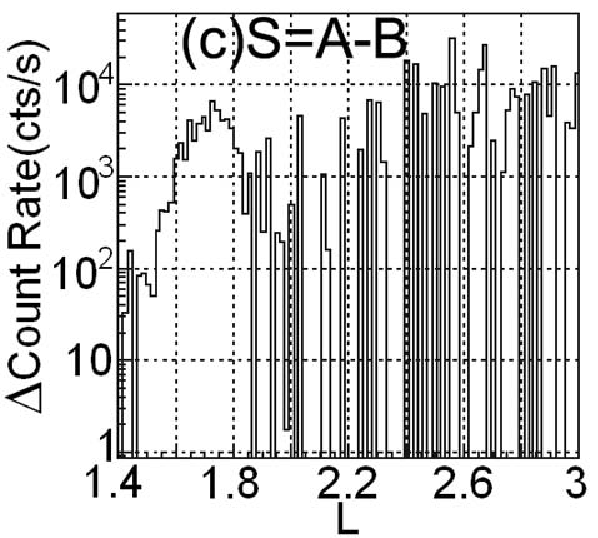}
\noindent\includegraphics[width=8pc]{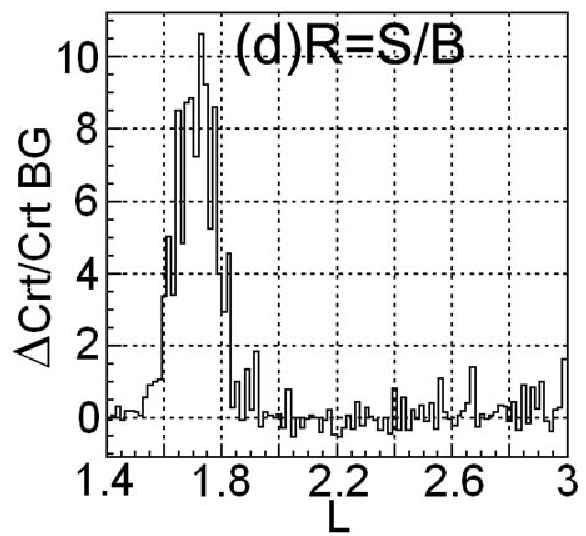}
\caption{The electron count rate distribution vs. L in the area of
NL. Longitude: $105^{\circ}\sim170^{\circ}$.\label{rangeL}} }
\end{figure}

\begin{figure}
\center{ \noindent\includegraphics[width=8pc]{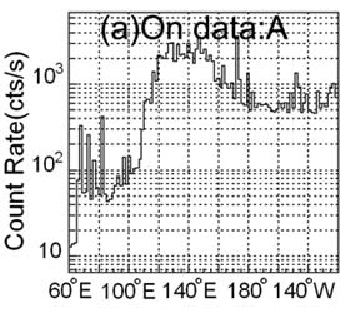}
\noindent\includegraphics[width=8pc]{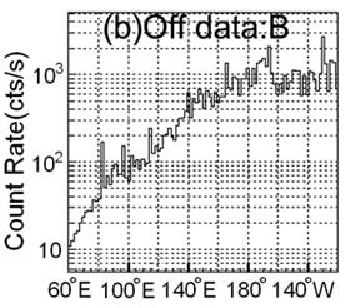}
\noindent\includegraphics[width=8pc]{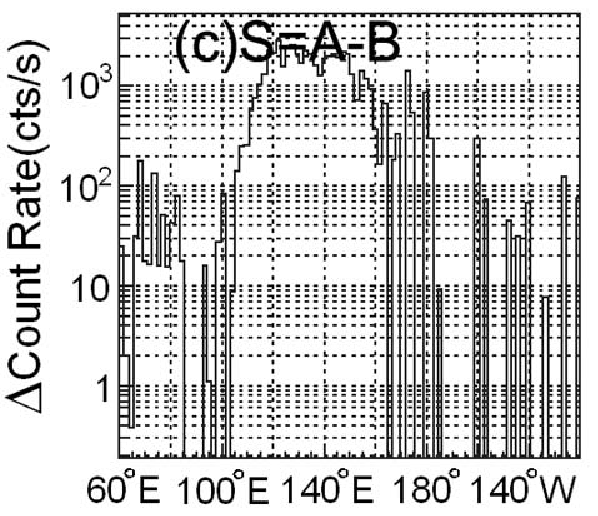}
\noindent\includegraphics[width=8pc]{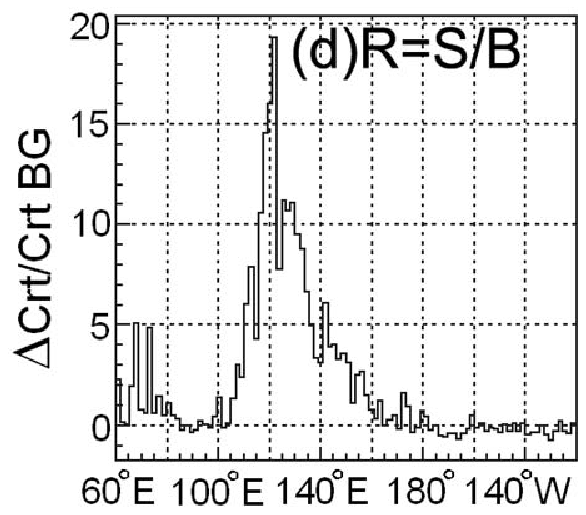}
\caption{The electron count rate distribution vs. geo-longitude in
the area of NL, L: 1.5$\sim$1.9.\label{rangelong}} }
\end{figure}

We selected the IDP data with electrons energy in 91$\sim$678keV
range, L value of (1.4$\sim$3) and geo-longitude value covering
180 degree. By (ON - OFF) method, the spatial distribution of NWC
electrons have been investigated quantitatively. As an example,
results for NL region are shown in Figure \ref{rangeL} and Figure
\ref{rangelong}. The average count rate of electrons were
calculated for whole given geo-longitude range in each value of
L$\sim$L+$\Delta$L. The results for ON data (A), OFF data (B),
signal (A-B), and the ratio of signal to background R (L) were
respectively presented in Figure \ref{rangeL}. Then The average
count rate of electrons were calculated for whole given L range in
each of geo-longitude interval $\lambda\sim\lambda+\Delta\lambda$,
the results for different condition including R ($\lambda$) were
presented in Figure \ref{rangelong}.

\begin{figure}
\center{
\noindent\includegraphics[width=6.5pc]{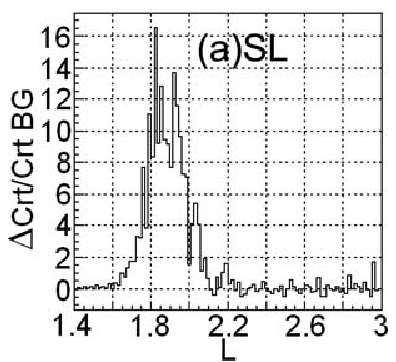}
\noindent\includegraphics[width=6.5pc]{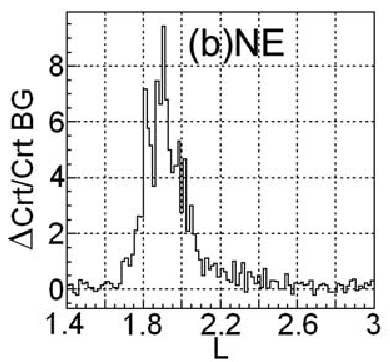}
\noindent\includegraphics[width=6.5pc]{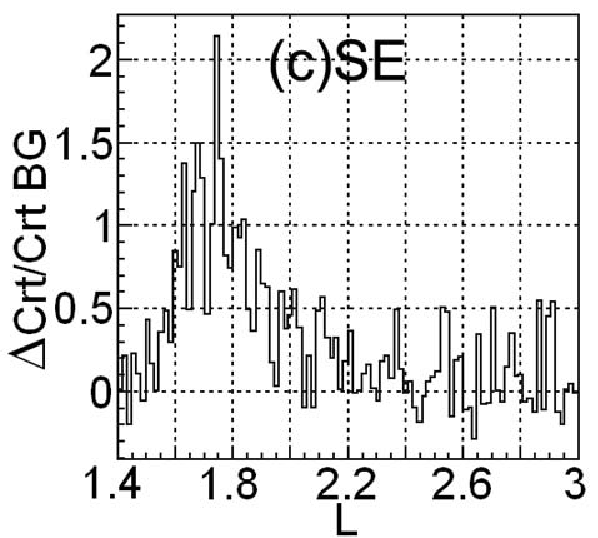}
\caption{Ratio of signal to background R(L) versus L-value in
other 3 regions. \label{rangeL3regions}} }
\end{figure}

\begin{figure}
\center{
\noindent\includegraphics[width=6.5pc]{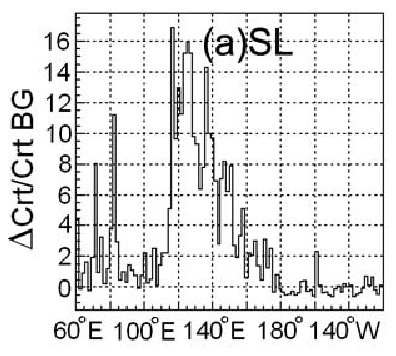}
\noindent\includegraphics[width=6.5pc]{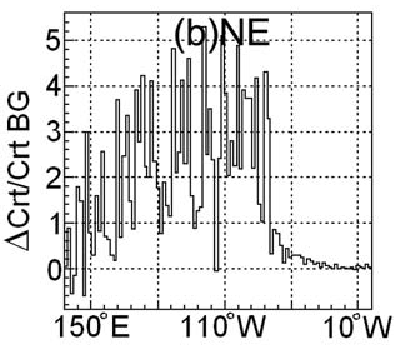}
\noindent\includegraphics[width=6.5pc]{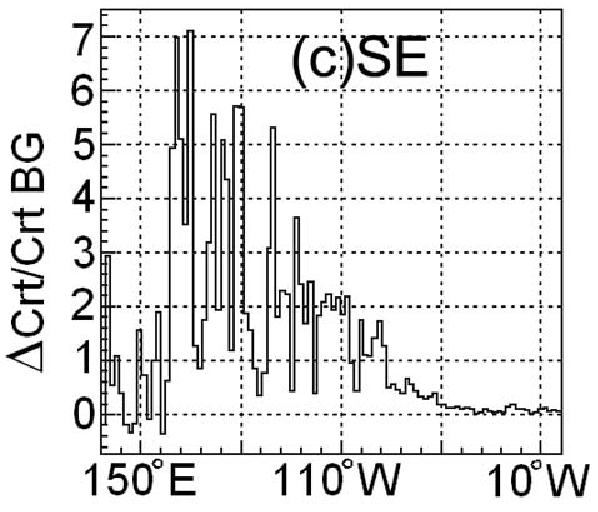}
\caption{Ratio of signal to background R(¦Ë) versus geo-longitude
in other 3 regions. \label{rangelong3regions}} }
\end{figure}

Using the same treatment, the analysis for the other three regions
of (SL, NE , SE) were also carried out. The results of R (L) and R
($\lambda$) for these regions are presented in Figure
\ref{rangeL3regions} and Figure \ref{rangelong3regions}.

It is clearly shown that the distributions of NWC electrons have
very complicated structures along the longitude and the magnetic
shell. In SE region£¬the effect from NWC VLF wave expands up to
300$^{\circ}$, which is deeply into SAA region due to the eastward
drift of NWC electrons. With the high background electron flux in
SAA, the NWC electrons still have S/B by factor of more than
1$\sim$2 times. The effect of expanding to the radiation belt
along with magnetic shell up to L$\sim$3 can also not to be
ignored. These sufficiently indicate that the VLF emission of NWC
transmitter powerfully influence the flux of the space electrons.

We define the range of drift lose cone for NWC electrons belts
according to R (L) and R ($\lambda$) distribution, and give the
name as "NWC electron belt", or "NWC belt" in brief, which are
listed in Table \ref{mainarea}.

\begin{table*}
\caption{The main area of the NWC belts(wisp
range).\label{mainarea}}
\begin{tabular}{c|cccc}
\hline
 &NL &SL &NE &SE\\
\hline\hline
L &1.5$\sim$1.9 &1.6$\sim$2.1 &1.6$\sim$2.2 &1.6$\sim$2.1\\
\hline
L@$R(L)_{max}$ &1.72 &1.82 &1.9 &1.75\\
\hline $\lambda(^{\circ})$ &105E$\sim$170E &110E$\sim$180E
&155E$\sim$70W
&150E$\sim$60W\\
\hline
$\lambda@R(\lambda)_{max}(^{\circ})$ &120E &117E &125E &160E\\
\hline
NWC local time &16:50$\sim$22:30 &16:50$\sim$22:30 &21:00$\sim$7:40AM &21:00$\sim$7:40AM\\
\hline
\end{tabular}
\end{table*}

It's worth mentioning that the data include both upward data
(night for the satellite) and downward data (day for the
satellite), so overlapping range exists along the longitude
direction. The longitude ranges in Table 1 correspond to the local
nighttime of the NWC station from 16:50 to 7:40am. Since the
increasing of the NWC electron flux can only be obviously observed
during the NWC night, the belts cover half the Earth with 180
degree along the longitude. The related ionosphere effect will be
discussed in the subsequent section \ref{sec4_2}.

\section{Energy spectrum of NWC electrons}
We investigated the energy spectrum of NWC electrons and its
variation in various situations. The spectra were accumulated by
On, Off, and (On-Off) respectively. The selected region is
basically according to Table \ref{mainarea}.
\subsection{Variation of energy spectrum along with L shell value}
According to the results shown in Section \ref{sec3}, we noted
that, at a certain altitude, the flux of NWC electrons varied with
L value of the satellite passed. The same effect exists for energy
spectrum. As an example, we presented the result for NL region in
Figure 9. The differential energy spectrum in three narrow range
of L shell are presented for ON, OFF, and (ON - OFF) status
respectively. We firstly revealed that: (a). Within NWC belt (L:
1.6 $\sim$ 1.65), flux of NWC electrons has a peak centered at
220keV with S/B of 35 times; (b) For higher L region of 1.9 $\sim$
1.95, the electron enhancement reduced and dominated in 300keV ~
1MeV, which reach to a maximum value of S/B about 1.3 times near
600keV; (c) For region in inner radiation belt L:2.05 $\sim$ 2.1,
a slot region is formed by the NWC electrons below 700keV,  with
S/B value reaching -60$\%$ near the peak of 300keV. In all three
situations, no NWC particles exist with energy above 1MeV except
the statistical fluctuation.

\begin{figure}
\center{
 \noindent\includegraphics[width=20pc]{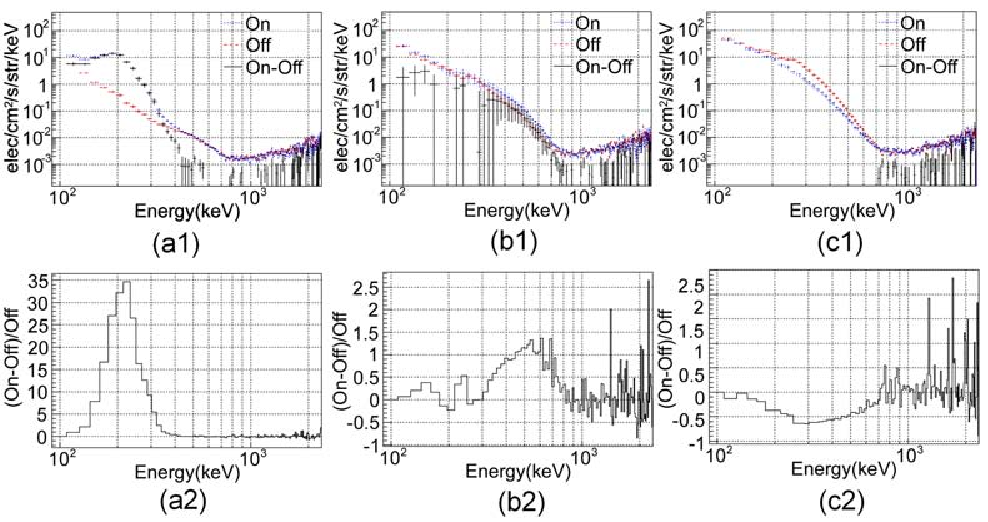}
 \caption{In NL region, the energy spectrum in three cases of On, Off,
 On/off in L range of (a), 1.6~1.65, (b), 1.9~1.95, (c), 2.05~2.1;
 and the ratio of Signal to background calculated by R=(On-Off)/Off
 and present in (a2), (b2), (c2). \label{totalSig_BG_NL}
 }
}
 \end{figure}

\begin{figure*}
\center{
 \noindent\includegraphics[width=30pc]{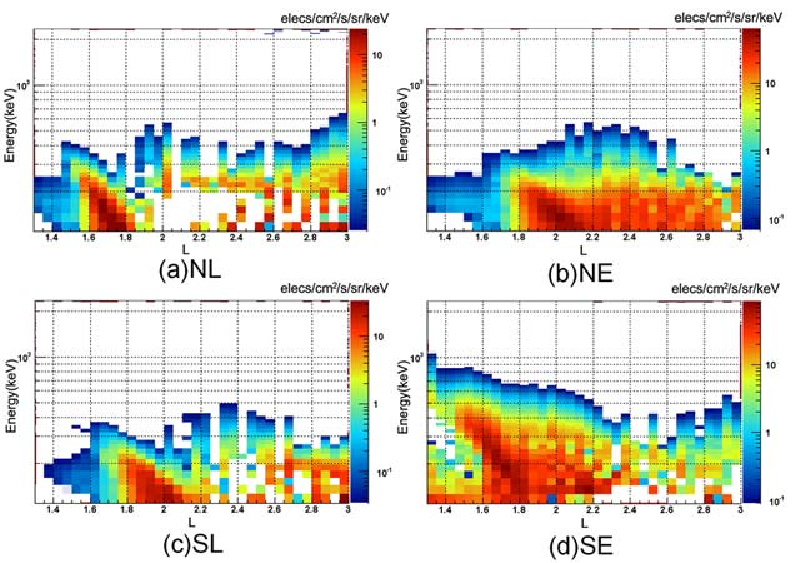}
 \caption{Spectrogram of NWC electrons along with L value in four different regions.
( The data with statistical fluctuation only has been ignored in
the figures.)\label{spectotal_pos}
 }
}
\end{figure*}

\begin{figure*}
\center{
 \noindent\includegraphics[width=30pc]{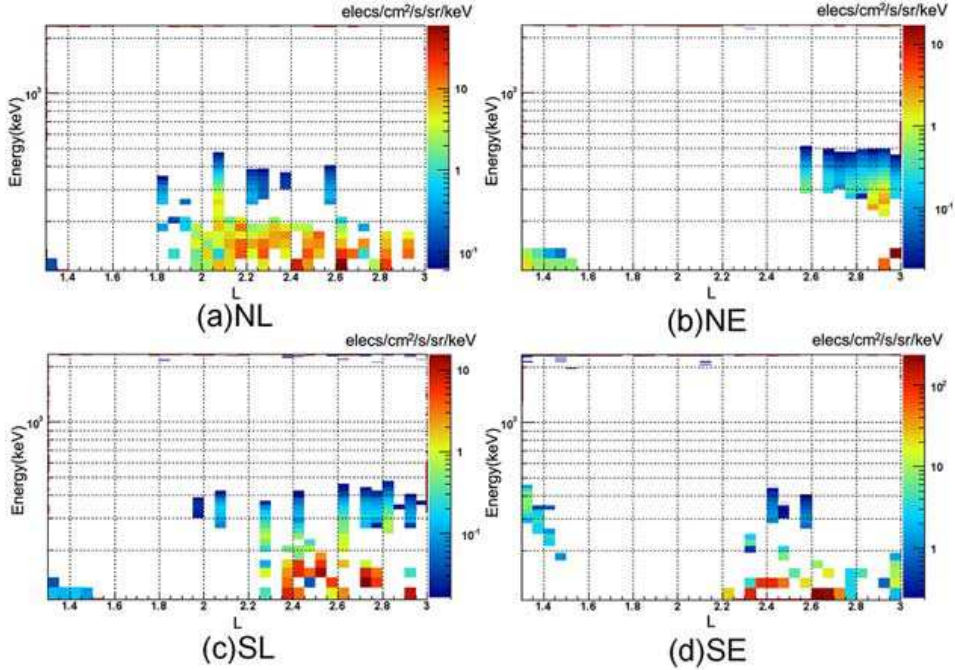}
 \caption{ Spectrogram of NWC losing electrons along with L value in four different regions
( The data with statistical fluctuation only has been ignored in
the figures.)\label{spectotal_neg}
 }
}
\end{figure*}
In order to accurately investigate the evolvement of NWC electrons
energy spectrum along with L shell value, we performed further
study of spectrogram for the L shell with value from 1.3 to 3 in
step size of 0.05 for the four NWC electrons belts, which have the
same longitude range with table \ref{mainarea}. The obtained
results of enhanced electron fluxes from ON-OFF value are shown in
Figure \ref{spectotal_pos} and that of lost electrons shown in
Figure \ref{spectotal_neg}.

From Figure \ref{spectotal_pos} and \ref{spectotal_neg}, we can
find the following characteristics
of NWC electrons energy spectrum:\\
1), In all 4 regions with electrons precipitation, the spectra
show obvious "wing" structures in relative lower L shell range.
The main part of "wing" structures correspond to NWC electrons
belts£¬which have been defined in Table \ref{mainarea}.\\
2), In Local regions, there are obvious slot effect caused by
electron loss in the energy range of 100 $\sim$ 200keV. The
electron loss is most obvious in north region (NL) corresponding
to L shell value in 1.95 $\sim$ 2.5. The spectrum shows irregular
structure in south
region (SL) corresponding to inner radiation belt with L > 2.2.\\
3£©£¬In Eastward regions, there are no obvious slot effect of NWC
electrons. The enhancement in NE and SE region with energy <300keV
extend to L $\sim$ 2.5 and even to L $\sim$ 2.9 respectively. The
fluxes of lost electrons are rather lower, the distributions are
mainly in 300 $\sim$ 400keV range and slightly different between
in SE and in NE.

Related to the physical mechanism, our understanding is that, the
"wing" structures should be most likely caused by local
interactions between original 19.8 kHz of VLF wave and space
electrons, which induce the electrons pitch angle scattering and
precipitation into loss cones. The original VLF wave, most
probably, come from NWC transmitter by direct ducted propagation.
And more electrons covered wider L shell range in Eastward are the
results from drift of precipitation electrons along with longitude
and magnetic shell, as well as from more electron resource
provided by SAA and inner radiation belt.

There are two more factors need to be mentioned here. First, our
analysis uses a geographic longitude coordinates rather than the
geomagnetic longitude coordinates. It may cause the difference
between north and south, east and west, for example, the covered L
range of "wing" feature are more similar between NL and SE, as
well between SL and NE. Another factor is the "east-west" effect
of electron fluxes caused by opposite pointing of IDP aperture
during upward and downward of the orbit. The difference is usually
equal to around 16$\%$ in mid-latitudes of northern hemisphere NL
(0 $\sim$ 28$^{circ}$) (Li XQ et al., 2010). The effect for NWC
belts in higher latitude can be roughly estimated from Figure
\ref{meansigmacurve}. The differences of background fluxes between
two regions could reach to two times or more. Similar effect for
NWC electrons can also be seen in the figure.

\subsection{Differential energy spectrum of NWC electrons and "Day-Night" effect}\label{sec4_2}
\begin{figure}
\center{
\noindent\includegraphics[width=20pc]{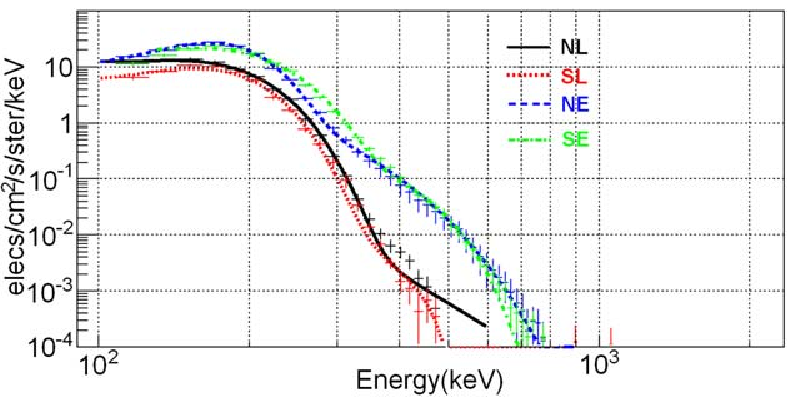}
\noindent\includegraphics[width=20pc]{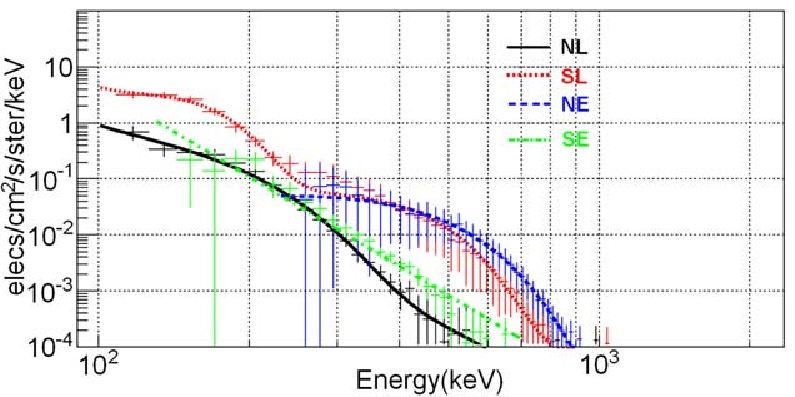}
\caption{Differential energy spectrum S(E) of NWC electrons in
drift loss cone. (a): During NWC night time, (b):During NWC day
time. \label{idpspecnightday}} }
\end{figure}

By (On-Off) method, we obtained total differential energy spectra
of NWC electrons, which are accumulated from each of 4 regions
defined in Table \ref{mainarea}. Figure \ref{idpspecnightday}(a)
presents the spectra related to NWC local night time using upward
data for SL and NL region, and downward data for SE and NE region.
Figure \ref{idpspecnightday}(b) is the spectra related to NWC
local daytime using downward data for SL and NL region, and upward
data for SE and NE region. For each spectrum, the average energy,
the integrated energy density and the integrated flux density, as
well as the differences of these physical parameters between NWC
day time and night time have been derived and listed in Table
\ref{CharacNight} and Table \ref{CharacDay}, respectively.

\begin{table*}
\caption{Characteristics of energy spectrum of NWC electron belts
during NWC nighttime, the values in parenthesis are
errors.\label{CharacNight}}
\begin{tabular}{c|ccc}
\hline
Area &Average Energy &Integrate influence &Integrated Flux\\
 &(keV) &($10^{-4} erg/cm^{2}/s/sr$) &($electrons/ cm^{2}/s/sr$)\\
\hline\hline
NL &167.0 (11.1) &3.59 (0.12) &1344.1 (43.7)\\
\hline
SL &169.8 (12.8) &2.68 (0.10) &986.8 (36.8)\\
\hline
NE &176.1 (7.4) &7.60 (0.16) &2696.2 (54.4)\\
\hline
SE &185.2 (8.5) &7.54 (0.17) &2543.5 (58.0)\\
\hline
\end{tabular}
\end{table*}

\begin{table*}
\caption{Characteristics of energy spectrum of NWC electron belts
(during NWC daytime), the values in parenthesis are
errors.\label{CharacDay}}
\begin{tabular}{c|cccc}
\hline
Area &Average Energy &Integrate influence &Integrated Flux &Flux ratio\\
 &(keV) &($10^{-4} erg/cm^{2}/s/sr$) &($electrons/ cm^{2}/s/sr$) &(night/day)\\
\hline\hline
NL &157.7 (39.7) &0.096 (0.012) &38.0 (4.8) &35.4\\
\hline
SL &161.0 (31.7) &0.601 (0.065) &233.7 (20.8) &4.2\\
\hline
NE &378.6 &0.071 (0.032) &11.7 (5.9) &231.1\\
\hline
SE &203.8 (109.5) &0.063 (0.016) &19.4 (5.4) &130.9\\
\hline
\end{tabular}
\end{table*}

\begin{figure}
\center{
 \noindent\includegraphics[width=13pc]{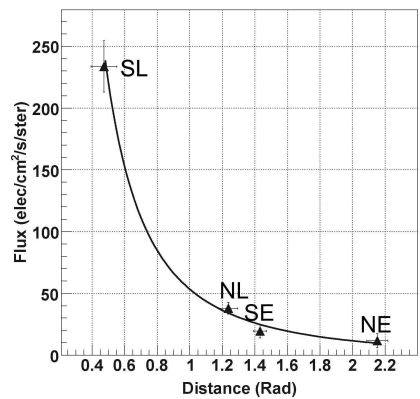}
 \caption{The relation between the electron flux and the angular distance to NWC. \label{Distance_Flux_fit}
 }
}
 \end{figure}

Here we give brief discussion about the result.\\
1), During NWC local night time, spectrum feature are different
between local and eastward region. The spectrum in SL and NL
region has almost the same feature, with the same peak below
~200keV and the same cutoff energy 5 at $\sim$ 500keV. There is
enhancement of electrons flux in Eastward region and the cutoff
energy increase to 800keV. This phenomenon indicates the existence
of continuous process of the electron precipitation and being
accelerated during drift. \\
2), During NWC Local daytime, we still can see the NWC electrons,
but the fluxes of electrons are rather weak, and the spectrum
features are quite different with each other. The cutoff energy of
NL and SL region are increased to 600keV and 700keV respectively.
The energy spectrum takes on anti-symmetric structure, the flux in
SL region is strongest and NL is weakest. The spectra in eastward
region show more absorption of electrons on their way of drift, so
that only the Electrons with E $>$ 250keV in NE region can be
observed. As we know that the average energy of wave-particle
resonance is $\sim$ 170 keV for 19.8 kHz VLF wave, so it at least
implicates that the original 19.8 kHz of VLF wave almost can not
reach to NE region due to ionosphere absorption during daytime.\\
3), We find out that the integrated electrons fluxes in daytime
observation can be well fitted by a quadratic function versus the
angle distance of D, which are calculated from coordinate of
central position for each belt related to the coordinate of NWC
ground station. The result is presented in figure
\ref{Distance_Flux_fit}, and formula \ref{fitformula}£¬where D
denotes the angle distance with unit of radian.

\begin{equation}\label{fitformula}
F=\frac{55.68}{D^{2}}-2.37 electrons/cm^{2}/s/sr
 \end{equation}

It provided the evidence that the wisp part of NWC electrons are
induced by original 19.8 kHz of VLF emission which are directly
propagated from NWC ground station,  It play major contributions
to NWC electrons during NWC night time, and the strong ionosphere
absorption during NWC day time. This is consistent with what we
discussed in the previous subsection.

\section{Verification of wave-particle theory from the pitch angle distribution}
IDP has no ability to identify the pitch angle of incident
particles. So, the direction between the axis of IDP aperture and
the local magnetic field is taken as the pitch angle and recorded
into the database. For each of four NWC electron belts, we make
such kind of pitch angle distribution, as example, the results for
NWC night time in NL and SL region are presented in Figure 14. The
actual distribution has to be expanded by a normalized Gaussian
function with 32$^{\circ}$ FOV, here we take 16$^{\circ}$ as
3$\sigma$, to obtain the pitch angle response function of the
observational incident electrons. Analysis shows that the
distributions are quite different in different belt regions, as
well as between NWC day and NWC night. But in all of the results,
measurable electron pitch angle are distributes mostly within
60$^{\circ} \sim$ 110$^{\circ}$.

\begin{figure}
\center{
 \noindent\includegraphics[width=20pc]{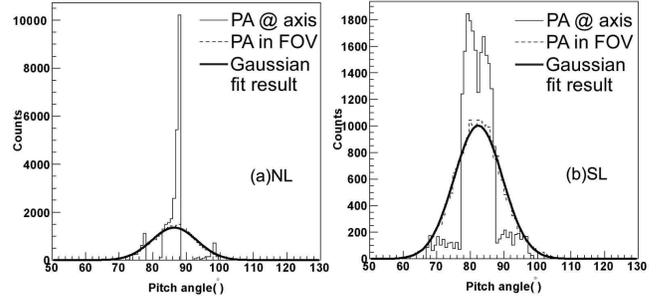}
 \caption{The pitch angle distribution of NWC electrons in NL and SL region during NWC night time. \label{PitchAngle}
 }
}
 \end{figure}

In the inner radiation belt, based on resonant interaction after
ignoring the energy diffusion£¬the quasi-linear diffusion equation
(Fokker-Planck equation) can be written as(\cite{Melrose1980}):

\begin{equation}\label{theoryformula1}
\frac{\partial f}{\partial
t}=\frac{1}{sin\alpha}\frac{\partial}{\partial
\alpha}(D_{\alpha\alpha}sin\alpha\frac{\partial f}{\partial
\alpha})
 \end{equation}

Where f($\alpha$,E,L) is density function in phase space, which
depends on local pitch angle of $\alpha$, kinetic energy of E and
Mc Ilwain parameter of L. $D_{\alpha\alpha}$ is local pitch angle
diffusion coefficient, which contains all information about
wave-particle interaction. As for NWC electrons with a given value
of L shell and kinetic energy of E, the value of local diffusion
coefficient $D_{\alpha\alpha}$  rely on the pitch angle
distribution.

Local diffusion coefficient of resonant interaction between
electrons and VLF wave can be written as \cite{Summers2005}:

\begin{equation}\label{theoryformula2}
D_{\alpha\alpha}=
\frac{\pi}{2}\frac{\Omega_{e}}{\rho}\frac{1}{(E+1)^{2}}\sum_{i}\frac{\Delta
b^{2}}{B^{2}}\frac{(1-\frac{x_{i}cos\alpha}{y_{i}\beta})^{2}|\frac{dx_{i}}{dy_{i}}|}{\delta
x|\beta
cos\alpha-\frac{dx_{i}}{dy_{i}}|}exp[-(\frac{x_{i}-x_{m}}{\delta
x})^{2}]
 \end{equation}

Where $\beta$= v/c, B is Earth's magnetic field; $x_{m}$ and
$\delta_{x}$ are reduced parameter defined as $x_{m}$
==$\omega_{m}$/$\Omega_{e}$, $\delta_{x}$
=$\delta$$\omega$/$\Omega_{e}$; $x_{i}$ and $y_{i}$ are reduced
variable with $x_{i}$ =$\omega_{i}$/$\Omega_{e}$, $y_{i}$
=c$k_{i}/\Omega_{e}$, respectively. Assuming pitch angle diffusion
is occurred only within equatorial plane and is valid for purely
field-aligned wave propagation, we obtain equatorial pitch angle
diffusion coefficient.

According to the DEMETER observation, to calculate equatorial
pitch angle diffusion coefficient, parameters are chosen as
follows: $\omega_{m}$ = 19.8kHz as frequency of maximum wave
power, $\delta$$\omega$= 150Hz and wave amplitude $\Delta$b =10pT.
Equatorial magnetic field is given by dipole model B = 3.11*
10$^{5}$/L$^{3}$T and equatorial plasma density N$_{0}$ = 880*
(2/L)$^{4}$cm$^{-3}$ which is within the range given by
(~\cite{Angerami1964}; ~\cite{Inan1984}).

\begin{figure}
\center{
 \noindent\includegraphics[width=16pc]{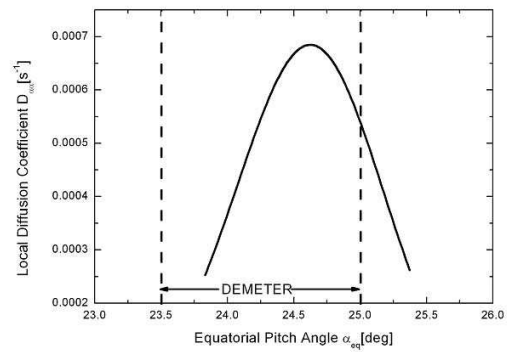}
 \caption{Local pitch angle diffusion coefficient as a function of
 equatorial pitch angle for given L and E. The area between two
 vertical lines corresponds to electrons with pitch angle ranging
 from 23.5$^{\circ}$ to 25$^{\circ}$ at equator which is 70$^{\circ}$
 to 90$^{\circ}$ at L=1.8 observed by DEMETER. Electron's kinetic energy E=220KeV.
\label{D_vs_PA_theory}
 }
}
 \end{figure}

From Figure \ref{D_vs_PA_theory}, we can clearly find that
electrons pitch angle is just within the FOV range or nearby of
IDP, where electrons participate the interaction with VLF wave
transmitted by NWC. As a result these electrons are observed by
DEMETER after diffusion process. The reason for the appearance of
"NWC man-made radiation belt" thus is probably explained by pitch
angle diffusion based on resonant interaction.

\section{Summary and discussion}
Based on DEMETER IDP observations at the altitude of 670 km, the
characteristics of the electrons precipitation, induced by VLF
wave emission from NWC station, have been firstly studied in
detail and quantitatively. With the long term observation data
related to NWC on and off, the background can be eliminated by
(On-Off) method with high statistics. For the first time, the
space distributions, the spectrogram via L shells, and the
differential energy spectra of NWC electrons have been obtained.

The results based on NWC night data indicate that, at 670 km
height, NWC VLF wave induced huge enhancement of electrons flux in
broad space range including the belts with "wing" structure. But
it also causes the electrons flux reduction which happened in
lower energy band and at some higher L regions. The electrons
precipitation, whatever enhancement or reduction, all can be
explained by pitch angle scattering of the electrons, which caused
by wave-particle resonance interactions, and possibly by the
electrons drift along the longitude. This phenomenon is a kind of
dynamic balanced process of the electrons flux variations. It
depends on which process is dominated for the electrons either
enhanced locally from higher L shell or moving down to the lower L
shell.

Used the data from both of NWC daytime and night time,
respectively, for the first time, the energy spectra are analyzed
for each of 4 belt regions. It has been shown that there are still
electrons enhancements during NWC daytime. But the fluxes are
rather weak comparing the results in NWC nighttime, even the
electrons in lower energy band are all disappear in eastward
region. It's already well known that this is caused by VLF wave
absorption of ionosphere at daytime. Interestingly, this effect
presents more regularity between the electron flux and the
distance of VLF wave propagations, which shown inverse-quadratic
function relation (Fig. \ref{Distance_Flux_fit} and equation
\ref{fitformula}). This easily links with the fact that the VLF
energy density attenuates by distance squared and then requires
the linearity relation between the flux of the precipitated
electrons and the energy density of the VLF wave. This analysis
provides reference for studying the ionosphere transmission
characteristics of VLF wave.

We may study more in further work.

%%% End of body of article:

%%%%%%%%%%%%%%%%%%%%%%%%%%%%%%%%
%% Optional Appendix goes here
%
%%%%%%%%%%%%%%%%%
% Geophysical Research Letters only allows an appendix without a letter.
%% You can get this result with
%  \section*{Appendix}
%  or
%  \section*{Appendix: Title}
%%%%%%%%%%%%%%%%%
%
% \appendix resets counters and redefines section heads
% but doesn't print anything.
% After typing  \appendix
%
% \section{Here Is Appendix Title}
% will print
% Appendix A: Here Is Appendix Title
%
% \section*{Appendix}
% will print
% Appendix
%
% \section*{Appendix: Here Is Appendix Title}
% will print
% Appendix: Here Is Appendix Title
%
% For only 1 appendix \appendix \section{Appendix} is preferred.
% which will print
% Appendix A

%%%%%%%%%%%%%%%%%%%%%%%%%%%%%%%%%%%%%%%%%%%%%%%%%%%%%%%%%%%%%%%%
%
% Optional Glossary or Notation section, goes here
%
%%%%%%%%%%%%%%
% Glossary only allowed in Reviews of Geophysics
% \section*{Glossary}
% \paragraph{Term}
% Term Definition here
%
%%%%%%%%%%%%%%
% Notation -- End each entry with a period.
% \begin{notation}
% Term & definition.\\
% Second Term & second definition.
% \end{notation}
%%%%%%%%%%%%%%%%%%%%%%%%%%%%%%%%%%%%%%%%%%%%%%%%%%%%%%%%%%%%%%%%
%
%  ACKNOWLEDGMENTS

\begin{acknowledgments}
This work is based on observations with the electric field
experiment ICE and the energetic particle experiment IDP embarked
on DEMETER, which is operated by the Centre National d'Etudes
Spatiales (CNES). The authors thank J. J. Berthelier, the PI of
ICE, and J.A. Sauvaud, the PI of IDP for the use of the data. The
authors would like to express their sincere thanks for Doctor
Zhenxia Zhang's help for this paper. This work was supported by
National High-tech R$\&$D Program of China (863 Program)
(2007AA12Z133).
\end{acknowledgments}

%% ------------------------------------------------------------------------ %%
%
%  REFERENCE LIST AND TEXT CITATIONS
%
% Either type in your references using
% \begin{thebibliography}{}
% \bibitem{}
% Text
% \end{thebibliography}
%
% Or,
%
% If you use BiBTeX for your References, please produce your .bbl
% file and copy the contents into your paper here.
%
% Follow these steps:
% 1. Run LaTeX on your LaTeX file.
%
% 2. Run BiBTeX on your LaTeX file.
%
% 3. Open the new .bbl file containing the reference list and
%   copy all the contents into your LaTeX file here.
%
% 4. Comment out the old \bibliographystyle and \bibliography commands.
%
% 5. Run LaTeX on your new file before submitting.
%
% AGU does not want a .bib or a .bbl file, but asks that you
% copy in the contents of your .bbl file here.

%Reference citation examples:

%...as shown by \textit{Kilby} [2008].
%...has been shown [\textit{Kilby et al.}, 2008].

%...as shown by \cite{jskilby}.
%...has been shown \citep{jskilbye}.

%% ------------------------------------------------------------------------ %%
%
%  END ARTICLE
%
%% ------------------------------------------------------------------------ %%

\end{article}

%% Enter Figures and Tables here:

% When submitting articles through the GEMS system:
% COMMENT OUT ANY COMMANDS THAT INCLUDE GRAPHICS.

% Figure captions go below this illustration; Table captions go above tables

% ONE-COLUMN figure/table, including eps graphics
%
% \begin{figure}
% \noindent\includegraphics[width=20pc]{chile-region-2010.eps}
% \caption{Caption text here}
% \end{figure}
% \end{document}
%
% \begin{table}
% \caption{}
% \end{table}
%
% ---------------
% TWO-COLUMN figure/table
%
% \begin{figure*}
% \noindent\includegraphics[width=39pc]{samplefigure.eps}
% \caption{Caption text here}
% \end{figure*}
%
% \begin{table*}
% \caption{Caption text here}
% \end{table*}
%
% see below for how to make landscape figures or tables

%%% End the article here:

\end{document}